**Spin-torque driven magnetic vortex self-oscillations in perpendicular magnetic fields**

G. Finocchio[1,*], V. S. Pribiag[2], L. Torres[3], R. A. Buhrman[2], B. Azzerboni[1]

[1]Dipartimento di Fisica della Materia e Ingegneria Elettronica, University of Messina, Salita Sperone 31, 98166 Messina, Italy.

[2]Cornell University, Ithaca, New York 14853-2501, USA.

[3]Departamento de Fisica Aplicada, University of Salamanca, Plaza de la Merced s/n, 37008 Salamanca, Spain

**ABSTRACT**

**We have employed complete micromagnetic simulations to analyze dc current driven self-oscillations of a vortex core in a spin-valve nanopillar in a perpendicular field by including the coupled effect of the spin-torque and the magnetostatic field computed self-consistently for the entire spin-valve. The vortex in the thicker nanomagnet moves along a quasi-elliptical trajectory that expands with applied current, resulting in "blue-shifting" of the frequency, while the magnetization of the thinner nanomagnet is non-uniform due to the bias current. The simulations explain the experimental magnetoresistance-field hysteresis loop and yield good agreement with the measured frequency vs. current behavior of this spin-torque vortex oscillator.**

The discovery that a spin-polarized current is able to change the magnetic configuration of a nanomagnet[1] has created technological opportunities for several types of nanoscale memories,[2] oscillators,[3] and radio frequency modulators and detectors.[4] In the case of the spin-torque driven auto-oscillator the magneto-resistance effect that produces the output signal can be related to several different mechanisms,[5,6,7] including: (i) the propagation of spin-waves, (ii) the excitation of bullet-modes, (iii) the excitation of spatially coherent and incoherent magnetization precession, and (iv) the excitation of precessional modes related to non uniform configurations. With respect to this latter case, recent experiments have shown that magnetic vortices can be brought into a self-oscillatory state in a spin-valve structure where the vortex is nucleated in the thicker nanomagnet.[8] A notable aspect of such spin-torque driven vortex dynamics is that the device can exhibit comparatively narrow linewidths (~ 300 KHz) in the GHz and sub-GHz frequency range, making such spin-torque vortex devices particularly interesting candidates for on-chip tunable microwave oscillators.

To advance the understanding of such devices here we report the results of advanced micromagnetic simulations of magnetic vortex dynamics in nanoscale spin-valve nanopillars that reproduce the steady-state precession of a vortex core with a quasi-elliptical trajectory. We take into account the full coupling of the two magnetic layers due to both magnetostatic interactions and spin-torque. We find that the interlayer coupling plays a very important role in these devices, as indicated by comparison to previous simulations[8] where the oscillation amplitude decayed slowly over time due to a simplified modelling of the interlayer spin-torque coupling. For the case of a high out-of-plane external field, we find that while the steady-state vortex-dynamics are localized in the thicker Py-layer, with the vortex core moving in an elliptical trajectory, the thinner layer magnetization is also highly non-uniform at the current levels where spin-torque dynamics begin. We also find that the temperature and the implementation of the torque do not substantially influence the frequency of the excited mode. Finally, we compare numerical results with experimental data explaining the features of the magnetoresistance-field hysteresis loop and obtaining a good agreement of the frequency vs. current curve.

For this micromagnetic modelling study we used a device geometry similar to that in Ref[8] (Py (5nm)/Cu (40nm)/Py (60nm) of elliptical cross sectional area (160 nm x 75nm)). The dynamics were simulated by solving the Landau-Lifshitz-Gilbert-Slonczewski (LLGS) equation,[9] with the magnetostatic field being computed self-consistently for the entire spin-valve. The spin transfer torque effect in the thinner Py-layer is simulated according to the following equation:

$$\mathbf{T}(\mathbf{m_p},\mathbf{m_f}) = \frac{g\ |\mu_B| j\varepsilon(\mathbf{m_f},\mathbf{m_p})}{e\gamma_0\ M_S^2 L_F}\left(\mathbf{m_f}\times(\mathbf{m_f}\times\mathbf{m_p})-\alpha\mathbf{m_f}\times\mathbf{m_p}\right) \quad (1)$$

where $g$ is the gyromagnetic splitting factor, $\gamma_0$ is the gyromagnetic ratio, $\mu_B$ is the Bohr magneton, $j$ is the current density assumed to be spatially uniform over the entire device, $L_F$ and $M_S$ are the thickness and the saturation magnetization for the thinner Py-layer, $e$ is the electron charge. Here $\mathbf{m_f}$ and $\mathbf{m_p}$ are respectively, the magnetization of the thinner Py-layer and of the top layer of the thicker Py-layer, where the latter is used as the polarizer layer for the spin-torque computation (the cell used for the discretization is 5x5x5nm$^3$). $\varepsilon(\mathbf{m_f},\mathbf{m_p}) = 0.5P(\chi+1)/(2+\chi(1-\mathbf{m_p}\bullet\mathbf{m_f}))$ is the polarization function which characterizes the angular dependence of the spin torque term, $P$ is the current spin-polarization factor and $\chi$ is the giant-magneto-resistance asymmetry parameter.

The spin transfer torque effect in the thicker Py layer is simulated according to:

$$\mathbf{T}(\mathbf{m_p},\mathbf{m_f}) = -\frac{g|\mu_B| j\varepsilon(\mathbf{m_p},\mathbf{m_f})}{e\gamma_0 M_{SP}^2 L_P}\left(\mathbf{m_p}\times(\mathbf{m_p}\times\mathbf{m_f})-\alpha\mathbf{m_f}\times\mathbf{m_p}\right) \quad (2)$$

where $\varepsilon(\mathbf{m_p},\mathbf{m_f}) = \varepsilon(\mathbf{m_f},\mathbf{m_p})$, where $\mathbf{m_f}$ is now the polarizer layer, $M_{SP}$ is the saturation magnetization for the thicker Py-layer, $L_P$ is the thickness over which the spin torque effect is exerted in the thicker Py layer and because the spin diffusion length in the Py is ~ 5nm,[10] we set this to be the thickness of the top discretized layer, i.e. the one closest to the thin Py layer. The magneto-resistance signal is computed over all ballistic channels as $r(\mathbf{m_p},\mathbf{m_f}) = \frac{1}{N_f} \sum_{i=1...N_f} r_i(\mathbf{m_{i,p}},\mathbf{m_{i,f}})$, where $N_f$ is the number of computational cell of the thinner layer and $r_i(\mathbf{m_{i,p}},\mathbf{m_{i,f}})$ is the magneto-resistance signal of the $i^{th}$ computational cell of the thinner layer ($\mathbf{m_{i,f}}$) computed with respect to the $i^{th}$ computational cell of the top discretized layer of the thicker layer ($\mathbf{m_{i,p}}$) by using a cosine angular dependence $r_i(\mathbf{m_{i,p}},\mathbf{m_{i,f}}) = 0.5[1-\cos(\theta_i)]$ ($\cos(\theta_i) = \mathbf{m_{i,p}} \bullet \mathbf{m_{i,f}}$).

For the simulations we employ a Cartesian coordinate system where the x-axis is the easy axis of the ellipse and the y-axis is the hard in-plane axis. By convention, positive current polarity corresponds to electron flow from the thinner to the thicker layer (+z-axis) of the spin valve, and we use: $M_S = M_{SP} = 650$ kA/m, $\chi = 1.5$ and $P = 0.38$, an exchange constant $A = 1.3 \times 10^{-11}$ J/m, and a damping parameter α=0.01. For the static case of no spin-torque current we simulated the magnetic behaviour of the spin-valves by solving the Brown equation ($\mathbf{m} \times \mathbf{h}_{eff} = 0$), with a residual of ≤$10^{-7}$ considered to be sufficiently low.

We simulated the magnetic hysteresis loop (Fig. 1(a) bottom) of the device structure, and qualitatively captured the major features of the giant magneto-resistance behavior as measured experimentally for near zero current bias (Fig. 1(a) top). In particular, above point A (above applied fields larger than 200mT) the vortex polarity is +1 (parallel to the field). As the field is swept to negative values, the polarity of the vortex core switches to -1 at point B ($\mu_0 H = $ -200mT). As the field is then swept back to positive values the core polarity switches back to +1 at point A ($\mu_0 H = $ 200mT). At zero field, point C the simulation shows the vortex core located near the center of the ellipse, with a small offset in the hard-in-plane axis due to the magnetostatic coupling with the thin Py-layer whose magnetization is aligned uniformly along +x or –x direction depending on the magnetic history. In the computation the field is tilted by 1° away from perpendicular direction along the +x direction to control the in-plane magnetization component. The magnetic configuration of the thinner Py-layer remains uniform for each value of field. The major quantitative difference between the simulated magnetic behavior and the experiment is that in the experimental loop, the switching fields of the vortex core polarity are ~ ±150mT, while the simulated values are somewhat larger, ~±200mT. This is most likely because thermal effects were not taken into account in the simulation of the near zero-current, magnetization loop. Possible shape imperfections due to device-to-device variations might also play a role.

We systematically studied the behaviour of this device structure under the application of a bias current sufficient to excite persistent dynamics, for a range of magnetic fields applied perpendicular to the plane of the thin film layers. In the case of a perpendicular field of 160mT (vortex polarity +1) the simulated vortex dynamics were characterized by a main excited mode with frequency in the range 1.8-2.2 GHz (Fig. 1(b) $T_{top}$ line). The frequency of this mode exhibits "blue shifting" as function of current, in agreement with experimental observations.[8] The simulated magnetization dynamics have a steady-state character, as can be observed from the temporal evolution of the average normalized magnetization of the thicker Py-layer ($<m_X(t)>$, $<m_Y(t)>$, Fig. 1(c), $J$=1.3 x $10^8$ A/cm$^2$), and the trajectory in the $<m_X>$-$<m_Y>$ plane (Fig.1(d)). The vortex core moves in counter-clockwise sense[11] with a quasi-elliptical trajectory.[12] The experimental "blue shift" functional dependence of the oscillation frequency on the out-of-plane field for a fixed current is also confirmed by our simulations which reveal that the average radius of the vortex trajectory increases with perpendicular field bias (not shown). For example for a current density of $J$=1.0 x $10^8$ A/cm$^2$, the average y-component of the magnetization oscillate from -0.08 to 0.08 at 100mT and from -0.15 to 0.15 at 160mT.

Fig. 2(a) show how the vortex positions (trajectories) differ, for the case of $\mu_0 H = 160$ mT, for two different bias current densities, $J = 1.0 \times 10^8$ A/cm$^2$ ('+') and $J = 2.0 \times 10^8$ A/cm$^2$ ('o') where we see that the average orbit also increases with current bias. Fig. 2(b) shows snapshots of the z-component of the magnetization of the top portion of thick layer at two different times during the vortex precessional period. The trajectories from this simulation of vortex dynamics in a uniform spin valve structure differ from those obtained previously for point-contact geometries in which the vortex moves into and out of the contact region, alternating its polarity and consequently the sense of rotation.[13] Our analysis of the simulation results indicate that the deviations from a pure elliptical trajectory are due to the strong non-uniform configuration in the thinner Py-layer, as shown, for example, in Fig.2(c) (left and right snapshots). In general, in the case of a significant field perpendicular to the plane, these non-uniform configurations are present even at the minimum current at which the vortex dynamics are excited. In contrast, at zero and near-zero applied field, above the onset current for vortex dynamics the magnetization of the thinner layer oscillates in a quasi-uniform configuration up to the current value that nucleates a second vortex in the thin layer and the GMR-signal then becomes the result of the relative motion of the two vortex cores (not shown).[14]

We studied the effect of several model parameters on the frequency of the vortex precession mode. We found no significant change in the dynamics if the torque is distributed across the whole thicker Py-layer or just to the top discretized section of this layer. As expected for a large magnetic volume system, we also find that including a thermal field[15] (T=300K, Fig.1(b) 300K line) does not affect the frequency of the dynamics, as illustrated by the two spectra in Fig. 3, which were computed via the micromagnetic spectral mapping technique,[16] for $J=1.3 \times 10^8$ A/cm$^2$. We do find a dependence of the frequency on the saturation magnetization of the thicker Py-layer, such that the frequency of the vortex dynamics decreases, and the onset current increases when increasing $M_{SP}$ to 800 kA/m, as shown in Fig.1(b). The frequency decrease with increasing saturation magnetization is in qualitative agreement with analytical results in point-contact geometries based on the rigid vortex model published in Ref.[17] for the case of a magnetic field applied perpendicular to the sample plane (see eq. 4 Ref. [17]).

Finally, we compare our numerical results with the experimental dynamical data (Fig. 1(b) of Ref. [8]). We determined the proportionality factor $\kappa$ between the experimental and the simulated currents using the same scaling procedure employed in Ref. [18], and obtained a quite reasonable value $\kappa$=0.65. The inset of Fig.3 shows a comparison between the experimental and computed frequency of the vortex self-oscillation at 300K and as can be observed, there is good agreement. Of course the linewidth computed numerically cannot be compared to the experimental data because the 50 ns simulation time limits the resolution to 20 MHz, while the experimental linewidths are usually well below 15 MHz.

In conclusion, our micromagnetic simulations reproduce the correct frequency dependence on current amplitude, a "blue shift," and show that: (i) the vortex core moves in the thicker Py-layer in an quasi-elliptical trajectory; (ii) the thinner Py-layer magnetization is in highly non-uniform configuration that changes dynamically as the vortex precesses about its trajectory; and (iii) that the mean radius of vortex orbit increases with the amplitude of an out-of-plane applied field. Our results also show that the combined spin-torque and magnetic coupling between the magnetic layers plays a crucial role in the explanation of the features of the magnetoresistance-field hysteresis loop and in obtaining persistent vortex precession. Given the importance of the coupling between the polarizer layer and the layer containing the vortex in determining the existence of the vortex precession it seems reasonable to surmise that the details of the precessional dynamics, e.g. linewidth, also depend critically upon this coupling.

This work was partially supported by Spanish Projects under Contracts No. SA025A08 and MAT2008-04706/NAN. This research was also supported in part by the Office of Naval Research/MURI program and by the National Science Foundation through the NSEC program support for the Center for Nanoscale Systems. Additional support was provided by NSF through use of the facilities of the Cornell Nanoscale Facility—NNIN and the facilities of the Cornell Center for Materials Research, an NSF MRSEC.

Figure 1 (color online): (a) experimental differential resistance (top) and simulated magnetoresistance (bottom) vs. out-of-plane field. (b) frequency of the vortex self-oscillation as function of current density ($\mu_0 H$=160mT) for different model parameters. The functional dependence are simulated with the spin transfer exerted on the top section only: T$_{top}$ ($T$=0K, $M_{SP}$=650 kA/cm$^2$), 300 K ($T$ = 300K, $M_{SP}$=650 kA/cm$^2$), and 800 kA/cm$^2$ ($T$ = 0K, $M_{SP}$ = 800 kA/cm$^2$). (c) Temporal evolution of the x and y component of the average normalized magnetization for $J$=1.3 x 10$^8$ A/cm$^2$, and (d) projection of the trajectory in the <m$_X$>-<m$_Y$> plane.

Figure 2 (color online): (a) spatial positions of the vortex core during its dynamics for two values of current densities $J$=1.0 x10$^8$ A/cm$^2$ ('+') and $J$=2.0 x10$^8$ A/cm$^2$ ('o'). (b) two snapshots of the vortex core position (the grayscale represents the amplitude of the z-component of the magnetization). (c) corresponding snapshots of the thinner Py-layer magnetization during the vortex dynamics in the thicker Py-layer. The arrows indicate the in-plane magnetization direction.

Figure 3 (color online): (main panel) spectra of the y-component of the magnetization computed by means of the micromagnetic spectral mapping technique ($J$=1.3 10$^8$ A/cm$^2$) for T=0K (black line) and T=300K (blue line). Inset: comparison between numerical and experimental data reported in Figure 1(b) of Ref[8] ($T$=300K, $M_S$=650 kA/cm$^2$).

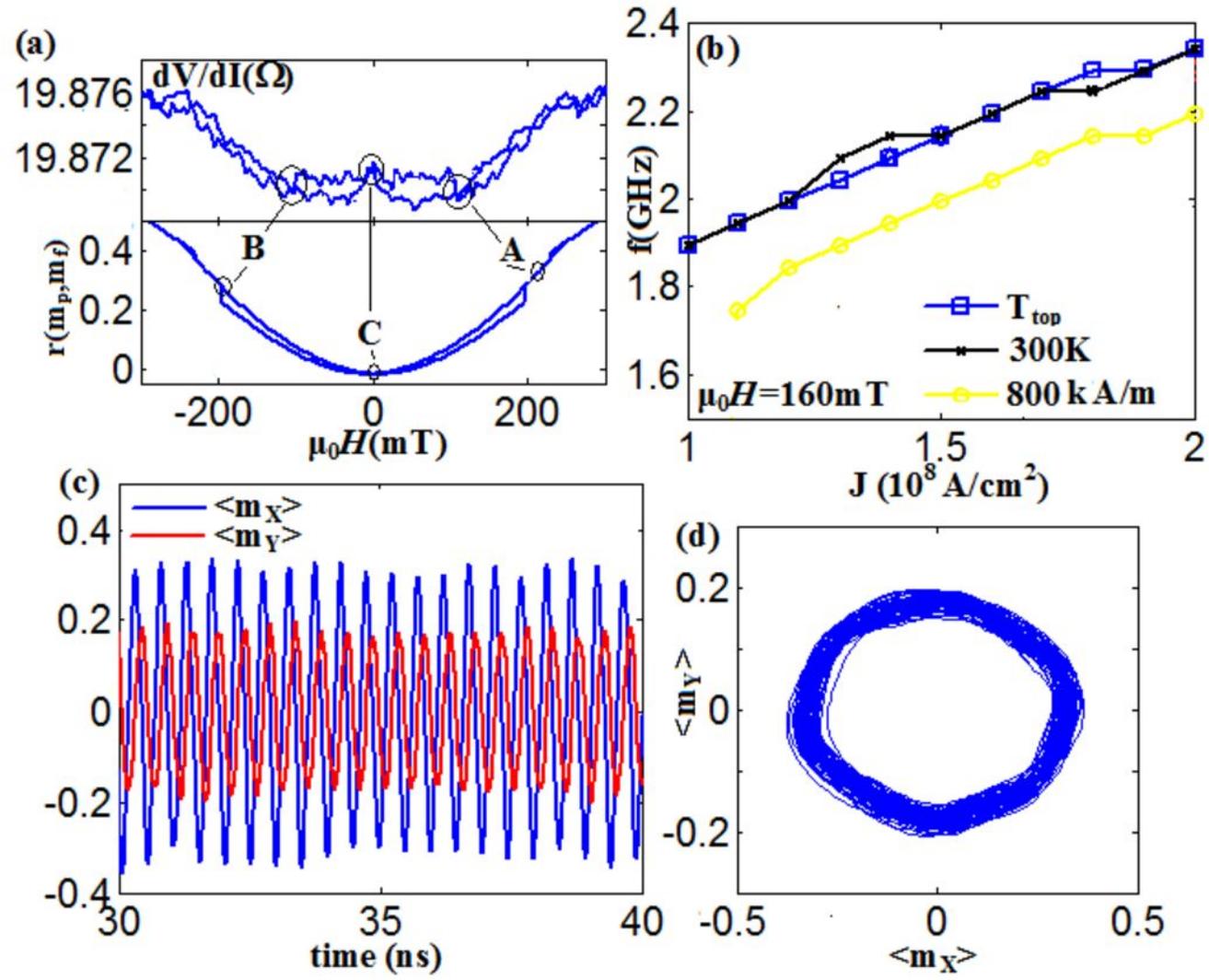

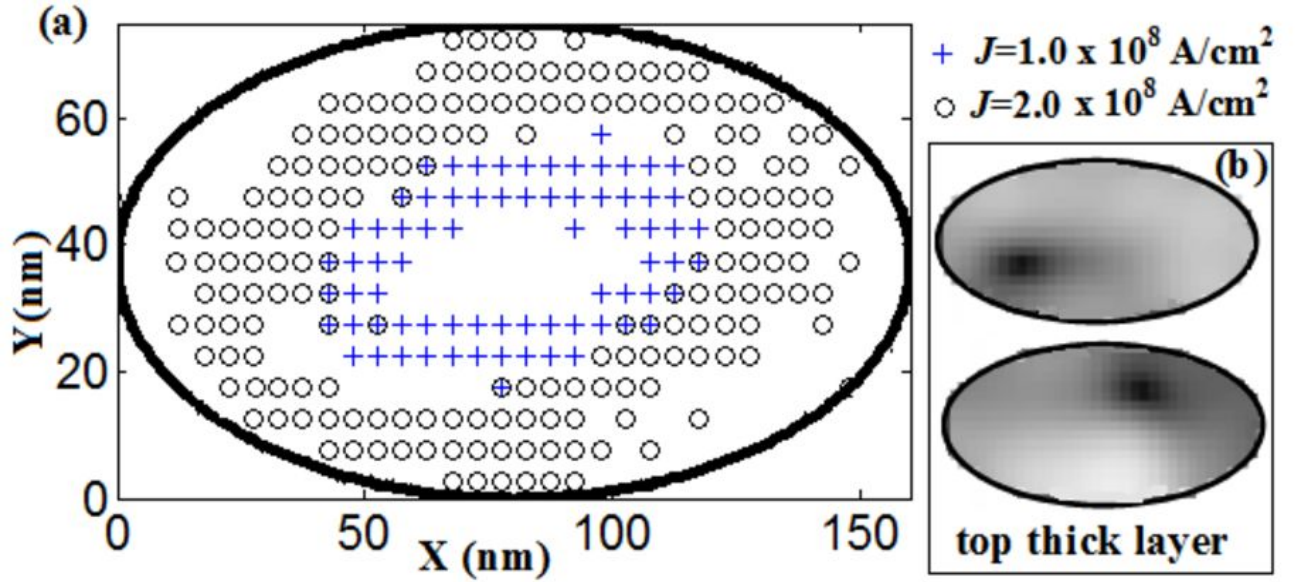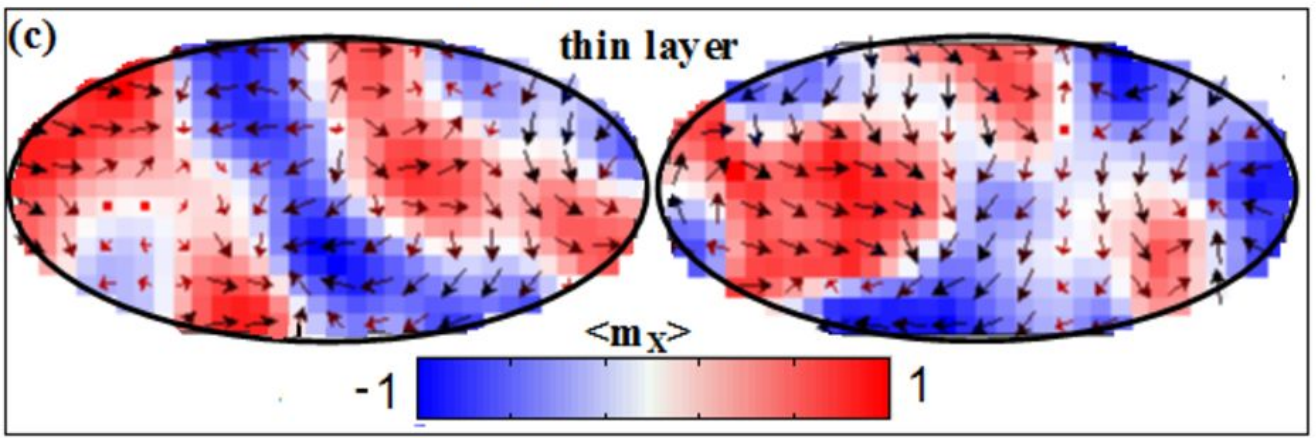

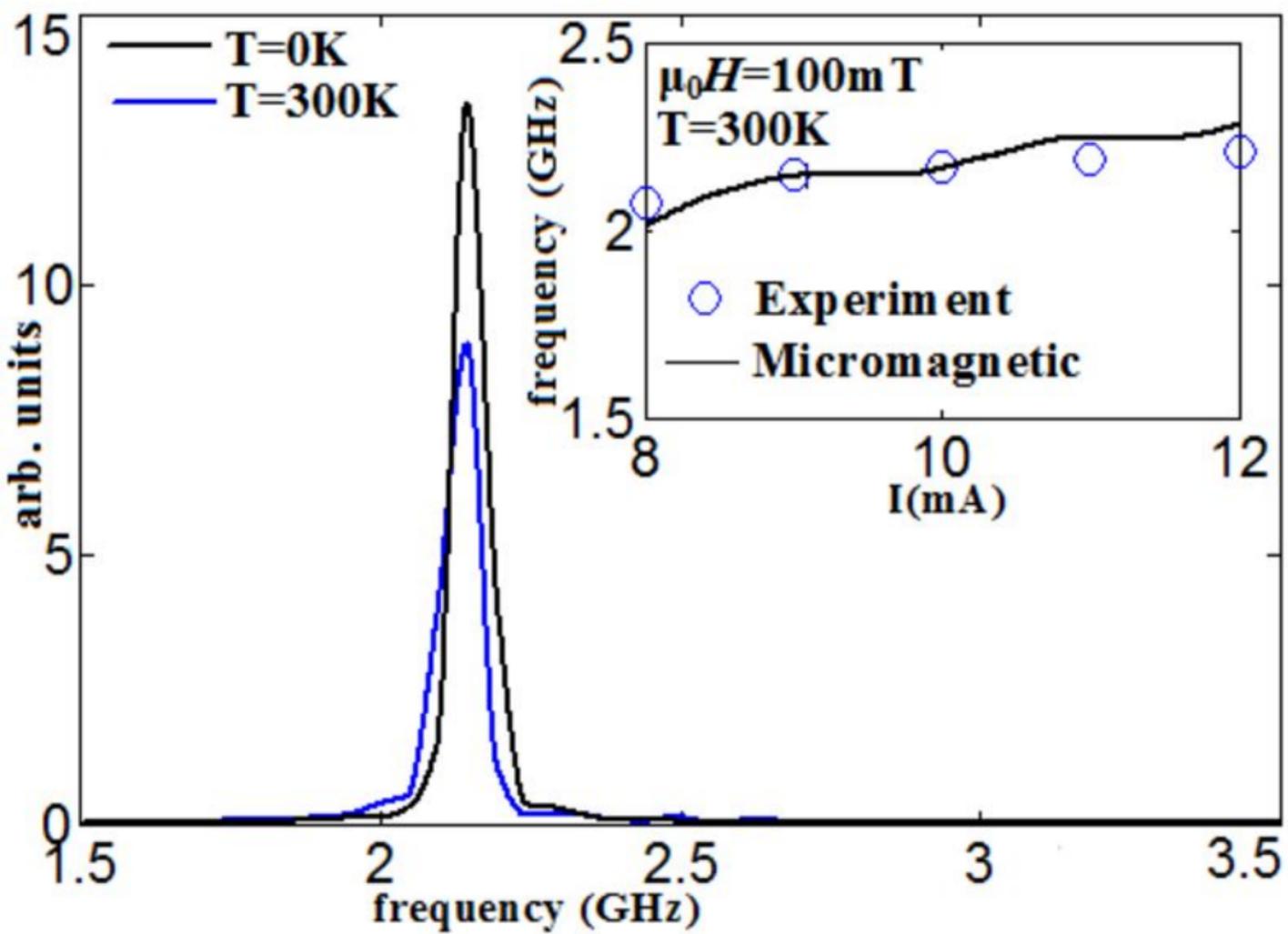